\documentclass[a4paper,11pt]{article}
\pdfoutput=1  % Ensures PDF output when compiling with pdflatex
\usepackage{jheppub} % JHEP package for formatting
\usepackage{graphicx} % Include figure files
\usepackage{dcolumn} % Align table columns on decimal point
\usepackage{bm} % Bold math
\usepackage{comment} % To comment out large sections
\usepackage{amsmath} % For advanced mathematical formatting
\usepackage{url} % For formatting URLs
\usepackage{color} % For colored text
\usepackage{hyperref} % For hyperlinks
\usepackage[utf8]{inputenc} % UTF-8 encoding
\usepackage{siunitx} % For SI units
\usepackage{mathtools} % For additional math tools
\numberwithin{equation}{section} % Number equations by section
 \author{Kartik Joshi \footnote{\href{mailto:kartikjoshi@iisermohali.ac.in}{Kartik Joshi}}}  
 \author{Satyajit Jena \footnote{\href{mailto:sjena@iisermohali.ac.in}{Satyajit Jena}}} 
\begin{document}

%\preprint{APS/123-QED}
%\preprint{JHEP-2024}

\title{Comparative Study of $\nu_{\tau}$ Event Numbers in INO and JUNO Detector from Bartol Flux}

%\author[a]{Kartik Joshi}
%\author[a]{Satyajit Jena}
\affiliation[a]{Department of Physical Sciences, IISER Mohali, Knowledge City, SAS Nagar, Sector 81, Punjab 140306, India}

%\emailAdd{kartikjoshi@iisermohali.ac.in}
%\emailAdd{sjena@iisermohali.ac.in}

\date{\today}
\maketitle
\flushbottom
\begin{abstract}

To expand our understanding of neutrino physics, scientific researchers of astroparticle physics direct their goal of detecting atmospheric $\nu_{\tau}$ in the GeV range. The effort will fundamentally unlock the nature of these elusive particles while also investigating the $\nu_{\mu}$ $\rightarrow$ $\nu_{\tau}$ oscillations. The Jiangmen Underground Neutrino Observatory (JUNO), which has already started its operations in 2024, and the India-based Neutrino Observatory (INO), which is not active right now but has future objectives in conducting research, have both emerged as key players in this field. These experiments used theoretical and experimental methodologies to understand the properties and behaviour of atmospheric $\nu_{\tau}$. The JUNO experiment, which has an estimated ability to detect around 50 events per year, and the INO, which used an impressive 50kTon iron slab as a detector, will contribute significantly in this domain [15]. The detection of all $\nu_{\tau}$ charged-current (CC) interactions with the detection material, which is divided into former and later events based on the timeline corresponding to scattering and capture in the detector, moreover, the KamLAND experiment is also capable of detecting these $\nu_{\tau}$ decays, however in smaller proportions, which could be possibly confused with background signals emerging from oscillations [16]. This has been studied for both experiments for $\nu_{\tau}$ nuclei cross-sections and $10^{-38}$ cm$^{-2}$ has taken a base for calculations Both INO and JUNO have $5\sigma$ sensitivity, which was exposed to 5 to 10 years.

\end{abstract}
 
\section{\label{sec:level1}Introduction}
%This sample document demonstrates proper use of REV\TeX~4.2 (and
%\LaTeXe) in mansucripts prepared for submission to APS
%journals. 
%Further information can be found in the REV\TeX~4.2
%documentation included in the distribution or available at
%\url{http://journals.aps.org/revtex/}.

The mathematical framework for neutrino oscillation in a vacuum as well as in matter is already constructed and given well-defined experimental results, which are proven in all neutrino oscillation experiments in various parts of the world, 
and the corresponding parameters such as $\theta_{13}$, $\theta_{23}$, and $\theta _{12}$ will be calculated together with mass hierarchy of $m^2_{1}-m^2_{2}$, $m^2_{2}-m^2_{3}$, and $m^2_{1}-m^2_{3}$.
The probability of vacuum oscillations is given by this: 
\begin{equation}
\begin{aligned}
 P(\nu_{\mu} \rightarrow{} \nu_{\tau}) = \sin^2\theta 
sin^2\frac{\Delta m^2L}{E}
\end{aligned}
\end{equation}
Down-going decay particles having channel $\pi$ $\rightarrow$ $\Bar{\mu}$ + $\nu_{\mu}$ have interaction with atmospheric nuclei, while up-going $\nu_{\mu}$ will oscillate to $\nu_{\tau}$ which can later interact with the detector to generate $\tau$ as an event, we can include the matter interaction with $\nu_{\mu}$ which can be converted to $\nu_{\tau}$ for all zenith angles [1].

\begin{equation}
\begin{aligned}
 P(\nu_{\mu} \rightarrow \nu_{\tau}) = 1.24\times\sin^2\theta_{m} 
sin^2\frac{\Delta m_{m}^2L}{E}
\end{aligned}
\end{equation}

Probability changes after interaction with earth's matter. The problem aims to find $\nu_{\tau}$ events detected by the detector at specific locations and all zenith angles. Their fluxes vary with energies, including earth's matter interaction. The study examines how their event probability increases in downward and upward scenarios. Analytical findings range from primary Super-K data sets. They have calculated fluxes for $\mu$ and $\nu_{\mu}$.

$\nu_{\tau}$ originate mainly from accelerated beams, high-energy supernovae, and extra-galactic sources at TeV ranges. This study helps us understand atmospheric sources. It explores if decaying pions and kaons lead to $\nu_{\mu}$ oscillations. Additionally, it analyzes how latitude and longitude affect neutrino fluxes. These variations are expected across different locations.

Since Bartol's collaborators have already estimated the variation of $\nu_{\mu}$ fluxes, we can use their theoretical calculations and estimations to calculate the $\nu_{\tau}$ fluxes at that specific location as calculated for $\nu_{\mu}$. One can also calculate the probability of $\nu_{\tau}$ events interacting with the detector at a specific time frame defined and calculated for the same. They depends on the detected precision specifications and locations. One can still predict the theoretical numbers coming out from calculations from Bartol fluxes by following the calculations and concepts highlighted in the paper. We have also incorporated the effects of neutrino's interaction with nuclei of iron via CC (charged current) and NC (neutral current) channels to calculate the outgoing energy of neutrino and subsequent scattering. This helps to understand its effects in neutrino event generation, taking products in zenith and azimuthal angles. Variation of fluxes of $\nu_{\mu}$ from Bartol fluxes is being calculated. Further matter oscillation effects will take care of tau fluxes for the earth profile. Latitude and longitudinal impact are taken care of for fluxes calculations. Numbers for $\nu_{\mu}$ around $10^8$ to $10^9$ are estimated with appropriate units across all zenith angles. This calculation uses Bartol flux from one of the decay channels $\pi$ $\rightarrow$ $\Bar{\mu}$ + $\nu_{\mu}$. Kaons are excluded due to their reduced branching ratio. A probability distribution used in oscillation parameters, including muon to $\nu_{\tau}$ conversion in the matter profile of earth around $4.3\text{g/cm}^3$. This incorporates the effects of matter interaction in the mixing angle and mass states of neutrinos. It also considers CC and NC processes, generating interactions with iron nucleons depending on neutrino energy. These interactions can fit into different classes, like elastic or quasi-elastic processes.

The idea is that the probability of getting $\nu_{\tau}$ events increases when we increase the baselines of oscillations. This requires tracking upward-going $\nu_{\mu}$ after their interaction with earth. The density of earth makes oscillation effects dominant and measurable in the detector. The flux of $\mu$$'$s  and $\nu_{\mu}$ in the decay channel of pions and kaons is generally calculated. This calculation is integrated over the spectrum of atmospheric depth. $\mu$$'$s lose energy in their corresponding decay. However, for neutrinos, the differential spectrum at a particular depth is the complete integration of the spectrum of production.

%\frac{dN_{\nu}(E_{\nu},X)}{dE_{\nu}} = $\[ \int_{0}^{X} {P_{\nu}}(E_{\nu},X,\theta),dX \]$
%\end{equation*}

The expression of Bartol flux can be converted into the required expression by assuming the upper limit to be infinite yielding the desired results. Cosmic rays, comprised of high-energy protons, interact with atmospheric nuclei to decay further into high-energy muons. These muons undergo secondary decay, with some decaying into electrons and anti-neutrinos, according to the number of decay channels defined in [3]. 
Branching ratios are also defined depending on their respective energies. The interaction of neutrinos with earth matter is also taken into account.
This increases the probability of getting $\nu_{\tau}$ events due to larger baselines compared to muons coming from the upper atmosphere. These baselines are approximately 30–40 km in 1D calculations [5]. After integrating each energy bin for all zenith angles defined in the primary datasets of Super-K, the muon neutrino numbers will be calculated using Bartol flux. 
This type of integral can be calculated at all zenith and energy bins by assuming the variables in the denominator as trigonometric functions.
\begin{equation}
\int_{0}^{E} \frac{E \cdot A_{\pi\nu}}{1 + B_{\pi\nu} \cos\theta \cdot E}\, dE
\end{equation}
$\\$
To find the number of $\nu_{\mu}$ for a particular energy bin, which has ranges of $0.112$ to $8.913$ GeV for all the datasets, one can calculate the above integral to know the actual approximate numbers theoretically. This calculation is quite unaffected by geomagnetic effects and altitude effects, which are also calculated for all the bins and have less than a $10^{-5}$ effect on the actual numbers calculated from theoretical values. Hence, they are considered negligible for most experiments conducted throughout the world at different latitudes and longitudes. On suitable assumptions, around $10^4$ high-energy protons in the energy ranges of GeV interact with atmospheric nuclei or molecules depending on the energy of protons in cosmic rays, After putting the value ranges of $A_{\pi_{\nu}}$ and $B_{\pi_{\nu}}$, which can take the value from 0 to 1[2], one can integrate the expression in required energy bins and later calculate for all the ranges of energy bins to find out the total numbers by adding all the values that come out from the integral. For all the different zenith angles, the approximate values will be calculated for all the theoretical cones built for a particular zenith angle in 3D and given in the table mentioned in the appendices. After accounting for all the values defined in [6], the Bartol flux is calculated as an integral, putting all the values in the expression.
\begin{equation}
\begin{aligned}
\int_{0.112}^{8.913} 10^4 \frac{1}{1+ 0.25E}\,dE 
\end{aligned}
\end{equation}
$\\$
Depending on the specific energy bins, the $\nu_{\mu}$ numbers will change from 50k to 100k, The expression is a reduced version of the actual atmospheric neutrino flux mentioned in [6] and can be calculated with specific parameters, also known as Bartol flux. This expression provides the results corresponding to the number of $\nu_{\mu}$ at specific zenith angles according to datasets or defined in the integral, which is from approximately 8.878$^{\circ}$ to 58.158$^{\circ}$ in intervals of zenith angle bins. This calculation is done from $cos \theta$ of the zenith angle from the primary dataset, and the suitable parameter has been calculated from the linear fitting function parameter, which is defined in each figure appropriately for the variables involved.
The integral can be solved by assuming a suitable trigonometric component for the terms involved in the denominator. The solution can be energy-dependent, ranging from $10^6$ to $10^9$ atmospheric $\nu_{\mu}$ events for typical zenith angles without any geomagnetic or other effects. Other factors like $A_{\pi_\nu}$ and $B_{\pi_\nu}$ will be taken care of for the Bartol flux defined ahead, and suitable $Z_{N_{N}}$ values will be taken to calculate the integral above.

\begin{itemize}
\item The (JUNO) is the largest observatory capable of detecting multiple particles either with FC (fully contained) events or through decay channels of secondary decays with a detector using scintillator techniques with liquids. At the same time, the INO in India has better cross-section measurement and a higher volume of detector interaction of 50,000 tons of active materials.

\item The project aims to precisely measure the ordering of the hierarchy of neutrino masses, with the detector having current 20,000-ton detector capabilities with around 77 photocells covered with high response times.

The achievement of 5$\sigma$ atmospheric tau $\nu_{\tau}$ events coming from matter interaction with the earth and its effects induced oscillations have been measured.

\item INO and JUNO use advanced software packages that maximize the precision required to achieve high-end results and measurements that a standard model can predict.

\item Significant contributions from Bartol fluxes will be considered to calculate the precise values of numbers of $\nu_{\tau}$ energies from theoretical models with parameters that will be taken care of, including branching ratios of $\nu_{\mu}/\nu_{\tau}$.

\item The ICAL detector has been built to study the different sources of neutrinos, including extragalactic, and it consists of 50k tons of iron frames that can be magnetized and used as RPC's as a current detector with 30k numbers of 1.8m $\times$ 1.8 size. Also, setting up an IICHEP centre for such endeavours for operation and maintenance will be taken care of from there so that technology can be applicable in other aspects.
\end{itemize}

\begin{figure}[h]
            \centering
            \includegraphics[width=0.40\textwidth]{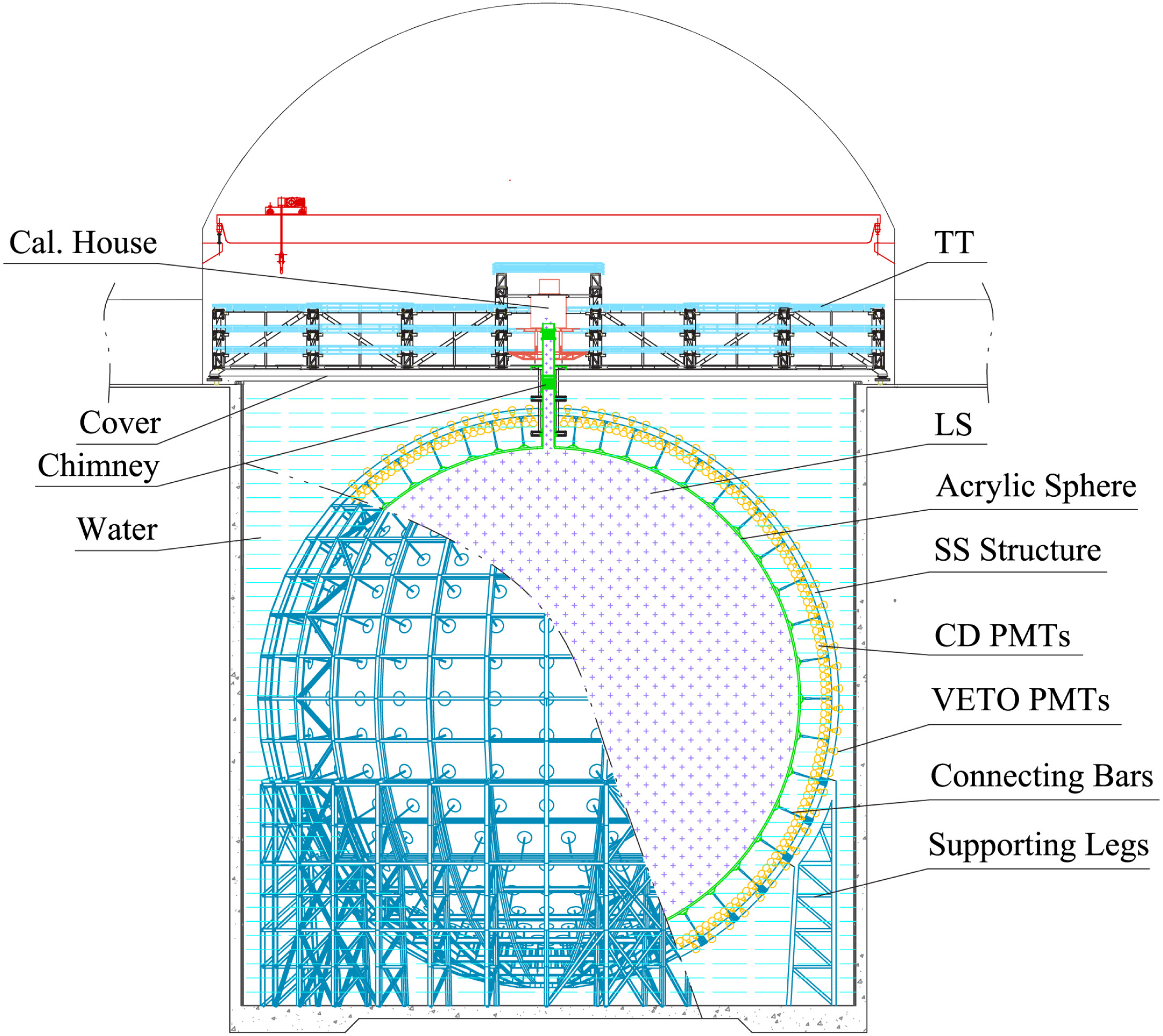}{[11]}
            \caption
%includegraphics[scale=0.55]{fitt-ZvsMuon.png}
%\includegraphics[scale=0.55]{zenith-energy.png}
{
JUNO detector has a 20,000-ton capacity with the possibility of $\nu_{\tau}$ events with high detector response, including other decay channels with $\tau$ [10].
}
\end{figure}

\begin{figure}[h]
            \centering \centering
            \includegraphics[width=0.30\textwidth]{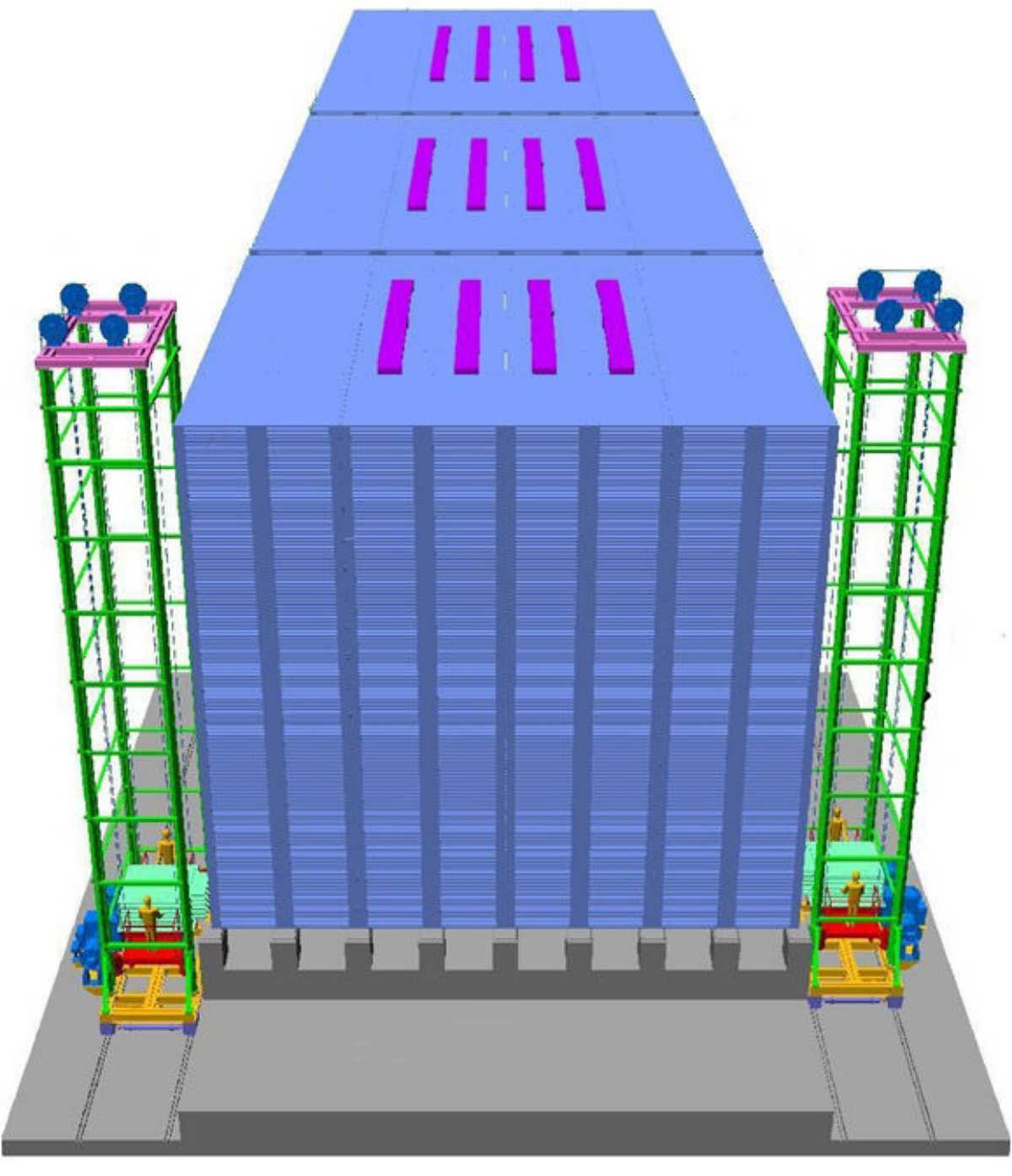}{[10]}
            \caption
{
INO detector has a 50,000-ton capacity, which enhances the possibility of more $\nu_{\tau}$ charged current events with a high detector response [9].
}
\end{figure}

There is still uncertainty in the atmospheric neutrino fluxes. We have used Bartol [2] flux. We expand the scope of this project to test other physics, including $\nu_{\tau}$ and $\nu_{\tau}{\Bar{}}$, using ICAL at INO. Several other detectors have been used to analyze $\nu{\tau}$ and anti-$\nu_{\tau}$ with corresponding decay channels in IceCube, KM3NeT, and worked on those energy ranges where geomagnetic effects will become significant.

\section{Discussion}

To calculate the $\nu_{\tau}$ flux from the Bartol $\nu_{\mu}$ flux, we start with the reciprocal of the Bartol flux expression in the standard probability flavour conversion formula. First, we integrate the $\nu_{\mu}$ flux to get muon numbers per unit area and unit time, obtaining the energy distribution. From this, we derive the probability distribution. We then determine the number of $\nu_{\tau}$ by multiplying the calculated $\nu_{\mu}$ numbers by the probability distribution of the conversion. We can include unit volume per year calculations depending on the detector configuration and its volume. NUANCE software can generate events and methods for analyzing $\pi \rightarrow \bar{\mu} + \nu_{\mu}$ channels. However, we have performed the statistical and analytical analysis for one decay channel. We can identify the longitudinal and latitudinal locations where $\nu_{\tau}$ fluxes can be measured. These measurements can be cross-checked with calculated $\nu_{\tau}$ values for every zenith angle across all energy bins, considering matter interactions. The Bartol flux for $\nu_{\mu}$ is given by:
\begin{equation}
\begin{aligned}
\frac{dN_\nu}{dE_\nu} = \frac{N_o(E_\nu)}{1 - Z_{NN}} \left[ \frac{A_{\pi \nu}}{1 + B_{\pi \nu} \cos\theta \frac{E_\nu}{\epsilon_\pi}} + 0.635 \frac{A_{K \nu}}{1 + B_{K \nu} \cos\theta \frac{E_\nu}{\epsilon_K}} \right]
\end{aligned}
\end{equation}
$\\$
Here, $1 - Z_{NN}$ represents the re-interaction of nucleons.
Integrating this expression gives the number of $\nu_{\mu}$ at a specific angle for all energy bins. The parameters $A_{\pi \nu}$ and $B_{\pi \nu}$ depend on the branching ratios, i.e., the probability of each decay process occurring. To simplify the problem, we consider the numbers from the decay of pions only, neglecting kaons. Thus, we have:
\begin{equation}
\begin{aligned}
\frac{dN_\nu}{dE_\nu} = \frac{N_o(E_\nu)}{1 - Z_{NN}} \frac{A_{\pi \nu}}{1 + B_{\pi \nu} \cos\theta \frac{E_\nu}{\epsilon_\pi}}
\end{aligned}
\end{equation}
$\\$
If the re-interaction is considered negligible in the analysis and $N_0(E_\nu)$ is a linear function of neutrino energy $E$, then the expression will become:
\begin{equation}
\begin{aligned}
\frac{dN_\nu}{dE_\nu} = E \times \frac{A_{\pi \mu}}{1 + B_{\pi \nu} \cos\theta \frac{E}{\epsilon}} dE d\theta    
\end{aligned}
\end{equation}
$\\$
Here, $A_{\pi \nu}$ and $B_{\pi \nu}$ are dependent on the decay channel and branching ratios. Their values lie theoretically in the ranges of 0 and 1 and depend on the fitting functional parameter $\gamma$. We have adopted a method for solving the integral, for example assuming the denominator equals some variable (t) and calculating further, then applying the limits in GeV ranges. These numbers are highly sensitive to energy in the GeV and MeV ranges, $\nu_{\mu}$ numbers come to be in the ranges of $10^9$ to $10^6$ at specific zenith angles. We can also find the distribution at all zenith angles for all energy bins. We can also calculate the number density of $\nu_{\tau}$ at specific locations, By multiplying the calculated $\nu_{\mu}$ values from the Bartol flux with the probability of conversion, we have considered matter effects with the earth's core density profile of $4.3 \text{ g/cm}^3$. After calculations, we find a conversion probability of approximately 0.688 from $\nu_{\mu}$ to $\nu_{\tau}$. Thus, the possible number of $\nu_{\tau}$ ranges from $10^4$ to $10^5$ per unit area and time, which might be detectable.
Given the GeV range of $\nu_{\mu}$, the possibility of quasi-elastic and weak scattering interactions is higher, which includes Matter-induced effects in neutrino oscillations having quasi-elastic and weak interactions with the iron nucleus.
\begin{itemize}
\item Similar calculations can be done using the NUANACE package that we calculated from Bartol fluxes. The probable values of $\nu_{\tau}$ come from oscillations and numbers of $\nu_{\mu}$ too, in the case of both the JUNO and INO detectors.
\item Around 85 hits is estimated in the ranges of 0–10 GeV for JUNO with charged currents of $\nu_{\tau}$ with $\nu_{\mu}$ as a background current.
\item The high energy $\nu_{\tau}$ charged current is measured with a $\tau$ lepton mass with a cut-off of 2.4 GeV, and corresponding high energy hadrons as pions are measured. A cut-off range of high energy oscillation events of $\nu_{\mu}$ to $\nu_{\tau}$ conversion has been recorded after matter interaction with the earth profile.
\item Fully contained events with deposited energy of 0.2 to 19 GeV can be differentiated from the background and precise resolution is achieved for $\nu_{\tau}$ events concerning $\nu_{\mu}$ events.
\end{itemize}
\section*{Step 1: Integrating \(\nu_{\mu}\) Bartol Flux}
The analysis of flux is given by the Bartol collaboration group known as Bartol \(\nu_{\mu}\) flux:
\begin{equation}
\begin{aligned}
\frac{dN_\nu}{dE_\nu} = \frac{N_o(E_\nu)}{1 - Z_{NN}} \left( \frac{A_{\pi \nu}}{1 + B_{\pi \nu} \cos\theta \frac{E_\nu}{\epsilon_\pi}} + 0.635 \frac{A_{K \nu}}{1 + B_{K \nu} \cos\theta \frac{E_\nu}{\epsilon_K}} \right)
\end{aligned}
\end{equation}

Where, \(1 - Z_{NN} \approx 1\), The re-interaction of nucleons is ignored.
We have focused on this $\pi$ $\rightarrow$ $\Bar{\mu}$ + $\nu_{\mu}$ decay channel due to its high branching ratio compared to K(kaon) $\rightarrow$ $\Bar{\mu}$ + $\nu_{\mu}$ decay channel:

\begin{equation}
\begin{aligned}
\frac{dN_{\nu_\mu}}{dE_\nu} \approx N_o(E_\nu) \frac{A_{\pi \nu}}{1 + B_{\pi \nu} \cos\theta \frac{E_\nu}{\epsilon_\pi}}
\end{aligned}
\end{equation}

\section*{Step 2: Integrating concerning Energy (GeV) and Zenith Angle ($\cos{\theta}$)}
We have to integrate the number of \(\nu_{\mu}\) for all zenith angles for all energy bins, Hence, the expression becomes:

\begin{equation}
\begin{aligned}
N_{\nu_\mu}(\theta) = \int_{E_{min}}^{E_{max}} N_o(E_\nu) \frac{A_{\pi \nu}}{1 + B_{\pi \nu} \cos\theta \frac{E_\nu}{\epsilon_\pi}} dE_\nu
\end{aligned}
\end{equation}

\section*{Step 3: $\nu_{\mu}$ $\rightarrow$ $\nu_{\tau}$ flavor oscillation probability}
This oscillation probability of $\nu_{\mu}$ $\rightarrow$ $\nu_{\tau}$ including the matter effects, is given by:

\begin{equation}
\begin{aligned}
P(\nu_{\mu} \rightarrow \nu_{\tau}) \approx \sin^2(2\theta_{23}) \sin^2\left( \frac{\Delta m^2_{32} L}{4E_\nu} \right)
\end{aligned}
\end{equation}

The profile of earth density is given by (average \(\rho = 4.3 \text{ g/cm}^3\)), after including the matter effect.

\section*{Step 4: \(\nu_{\tau}\) Flux Estimation}
Using the probability of conversion as below :  

\begin{equation}
\begin{aligned}
\frac{dN_{\nu_\tau}}{dE_\nu} = \frac{dN_{\nu_\mu}}{dE_\nu} \times P(\nu_{\mu} \rightarrow \nu_{\tau})
\end{aligned}
\end{equation}

After Substituting the integrated, we get \(\nu_{\mu}\) flux:

\begin{equation}
\begin{aligned}
N_{\nu_\tau}(\theta) = \int_{E_{min}}^{E_{max}} N_o(E_\nu) \frac{A_{\pi \nu}}{1 + B_{\pi \nu} \cos\theta \frac{E_\nu}{\epsilon_\pi}} \times P(\nu_{\mu} \rightarrow \nu_{\tau}) dE_\nu
\end{aligned}
\end{equation}

\section*{Step 5: Detector Considerations and Configuration}
For the detector under study(JUNO, INO), we have considered the volume and time to get number density:
\[
N_{\nu_\tau, \text{detected}} = N_{\nu_\tau} \times \text{detector efficiency} \times \text{detector volume} \times \text{time}
\]
\section*{Step 6: Detector Dependent Calculations}
For JUNO and INO(0-10 GeV):
\[
\text{Hits per year} \approx \text{number of } \nu_{\tau} \text{ unit area and time} \times \text{cross-sectional area of detector} \times 
\text{efficiency} \times \text{time}.
\]
\begin{comment}
\section*{Example Calculation}
\begin{enumerate}
    \item Integrate \(\frac{dN_{\nu_\mu}}{dE_\nu}\) over the relevant energy range (e.g., 0-10 GeV).
    \item Multiply by the conversion probability to get \(\nu_{\tau}\) flux.
    \item Consider detector specifics (e.g., JUNO’s and INO's volume and efficiency) to get the number of detected \(\nu_{\tau}\).
\end{enumerate}
\end{comment}
The equation [2.2] included the complete picture of the estimation done: 
\begin{equation}
\begin{aligned}
\int_{0}^{E} \frac{dN_{\nu_{\mu}}}{dE_{\nu_{\mu}}} \, dE \times \sigma \times \delta t \times N
\end{aligned}
\end{equation}
$\\$
\begin{comment}
$10^{4}$ (Approximately calculated numbers after matter interaction with earth as upgoing $\nu_{\tau}$ events from Bartol flux) $\times$ $10^{33}$ (Total number of nuclei interacted in 50k ton using unitary method with Avogadro's numbers definition) $\times$ $10^9$ (Total exposure time of total events in the detector in seconds for 10 years) $\times$ $10^{-38}$ (Interacted cross-section in $cm^2$ in usual units) $\times$ $10^{-13}$ (Lifetime of leptons here in our case it is $\tau$) $\times$ $10^{8}$ (Total cross-section area involved for 50k ton detector with $\nu_{\tau}$ numbers in $cm^{2}$ as defined in [10]). The whole calculation will give approximately 100 $\nu_{\tau}$ events for INO and 50 $\nu_{\tau}$ in the case of JUNO in a timeline of 10 years.

I am using the above analysis, the Bartol model to detect the actual numbers of $\nu_{\tau}$ having a cross-section in a timeline of 10 years with the inclusion of interaction with primary numbers of nuclei and a lifetime of $\tau$ leptons. Since INO has a 50,000-ton detector with $10m\times10m\times20m$, which is calculated to produce around 60 $\nu_{\tau}$ events, and JUNO has a 20,000-ton detector that can produce 50 $\nu_{\tau}$ with corresponding leptons per year.
\end{comment}

To calculate the number of $\nu_{\tau}$ events from the Bartol $\nu_{\mu}$ flux, we use the following parameters:

\begin{align*}
10^{4} & \quad \text{(Approximate $\nu_{\tau}$ events after matter interaction with earth, as upgoing events)} \\
\times 10^{33} & \quad \text{(Total number of nuclei interacted in 50,000 tons using unitary method)} \\
\times 10^{9} & \quad \text{(Total exposure time in the detector in seconds for 10 years)} \\
\times 10^{-38} & \quad \text{(Interaction cross-section in $cm^2$)} \\
\times 10^{-13} & \quad \text{(Lifetime of $\tau$ leptons)} \\
\times 10^{8} &   \quad \text{(Total cross-sectional area involved for a 50k ton detector in $cm^{2}$)}
\end{align*}

The overall calculation gives approximately 100 $\nu_{\tau}$ events for INO and 50 $\nu_{\tau}$ events for JUNO over 10 years.

Using this analysis with the Bartol model, we aim to detect the actual number of $\nu_{\tau}$ events. This includes the interaction with primary nuclei numbers and the lifetime of $\tau$ leptons. 

For the INO detector, the calculation predicts around 60 $\nu_{\tau}$ events over 10 years. The INO detector has a volume of $10\,m \times 10\,m \times 20\,m$, which is equivalent to 50,000 tons. For the JUNO detector, the calculation predicts about 50 $\nu_{\tau}$ events over the same period. The JUNO detector has a volume of 20,000 tons.

\subsection{\label{}Interpretation of Super K Data}
\begin{itemize}
    \item Preliminary datasets include error corrections, $\nu_{\mu}$ fluxes, zenith angles, azimuthal angles, and interaction depths with atmospheric nuclei. A total of 600 datasets are used to explore parameters for decay channel simulations and $\nu_{\mu} \rightarrow \nu_{\tau}$ oscillation analysis. We plotted the first 20 datasets for our analysis, as mentioned in [11]. We consider uncertainties due to detector response and statistical errors. Including all datasets would provide the complete area under the curve from the Super K experiment.
    \item Super K data, spanning 15 years, provides information on $\tau$ decay channels from $\nu_{\tau}$ sources. However, HyperK lacks sensitivity analysis. Neutrinos can originate from various sources, such as cosmic rays interacting with atmospheric nuclei. This study focuses on $\pi^{\pm}$ and heavy hadron fluxes [\ref{2}]. We are particularly interested in fluxes from $\nu_{\mu}$ oscillations in upward and downward going events.
    \item The JUNO detector analysis includes interaction cross-sections with iron nuclei and $\nu_{\mu} \rightarrow \nu_{\tau}$ oscillation events. Super K's extensive timeline and comprehensive datasets support this detailed study. We aim to understand $\tau$ decay channels and their significance in neutrino oscillation experiments. This enhances the overall understanding of neutrino interactions and behaviours.
    \item All the calculations mentioned in Table 1 to Table 5 have been calculated from the datasets\cite{14}. Estimation has been done for downgoing and upgoing $\nu_{\tau}$ events for all zenith angles with corresponding energy in 1-10 GeV ranges including matter interactions with different baselines before hitting the detector. The events calculated in the discussion section [2] are relevant when actual numbers hit the detector to generate actual events.
\end{itemize}

\begin{table}
    \centering
    \caption{First 20 primary data sets from 600 for Super K experiment in Japan.}
    \label{tab:your-label}
    \begin{tabular}{|c|c|c|}
        \hline
        Energy (GeV) & Zenith angle & Muon flux  \\
        \hline
        0.112	 &-0.95	 &  762.52932  \\
        0.141    &-0.95  &  709.47164  \\
        0.178    &-0.95  &  678.39415  \\
        0.224    &-0.95  &  572.28661  \\
        0.282    &-0.95  &  415.29105  \\
        0.355    &-0.95  &  335.71681  \\
        0.447    &-0.95  &  267.31772  \\
        0.562    &-0.95  &  202.00585  \\ 
        0.708    &-0.95  &  151.91063  \\
        0.891    &-0.95  &  114.68026  \\
        1.122    &-0.95  &  80.311611  \\
        1.413    &-0.95  &  55.380521  \\
        1.778    &-0.95  &  38.413084  \\
       2.239    &-0.95  &  27.045827  \\
        2.818    &-0.95  &  18.391713  \\
       3.548    &-0.95  &  11.799715  \\
        4.467    &-0.95  &  8.0611414  \\ 
        5.623    &-0.95  &  5.0495139  \\
        7.079    &-0.95  &  3.4698978  \\
        8.913    &-0.95  &  1.3658999  \\
    \hline
    \end{tabular}
\end{table}
%\end{comment}

\begin{table}
    \centering
    \caption{Calculations from primary data sets for Super K in Japan.}
    \label{tab:your-label}
    \begin{tabular}{|c|c|c|}
    \hline
    $\theta$	 & L$/$E   & $\nu_{\mu}$   \\
    \hline
54.4309905374282			&113767.857142857&           9116.12308104447\\
54.4309905374282			&90368.7943262411&           9096.23566651513\\
54.4309905374282			&71584.2696629214&           9083.97310571742\\
54.4309905374282			&56883.9285714286&           9068.55866050327\\
54.4309905374282			&45184.3971631206&           9049.23191873109\\
54.4309905374282			&35892.9577464789&           9024.99194373395\\
54.4309905374282			&28505.5928411633&           8994.87398421569\\
54.4309905374282			&22672.5978647687&           8956.92565085622\\
54.4309905374282			&17997.1751412429&           8909.81009651807\\
54.4309905374282			&14300.7856341198&           8851.03944539681\\
54.4309905374282			&11356.5062388592&           8778.09814593396\\
54.4309905374282			&9017.69285208776&           8688.29047623125\\
54.4309905374282			&7166.47919010124&           8577.45482038609\\
54.4309905374282			&5690.93345243412&           8442.19212634483\\
54.4309905374282			&4521.64655784244&           8277.61513434837\\
54.4309905374282			&3591.31905298768&           8079.33368469835\\
54.4309905374282			&2852.47369599284&           7843.01280107661\\
54.4309905374282			&2266.05015116486&           7564.33594288033\\
54.4309905374282		    &1799.97174742195&           7240.28620667676\\
54.4309905374282		    &1429.59721754747&           6907.15465146516\\
          
\hline
\end{tabular}
\end{table}

\begin{table}
    \centering
    \caption{Actual numbers of $\nu_{\tau}$ downwards and upwards after probability conversion with baselines of 30 km to 12742 km(diameter of the earth)}
    \label{tab:your-label}
    \begin{tabular}{|c|c|c|c|}
    \hline
$\nu_{\mu}$-$\nu_{\tau}$Upwards&$\nu_{\mu}$-$\nu_{\tau}$Downwards&$\nu_{\tau}$Upwards& $\nu_{\tau}$Downwards \\
    \hline
	
0.5097216921	&0.517653498932	&4645.1065&	4717.3892\\
0.2523904527	&0.356725094972	&2298.2985&	3248.3826\\
0.7856602752	&0.236573504521	&7146.5510&	2151.9283\\
0.1516331833	&0.154559219193	&1377.4317&	1404.0117\\
0.8820408670	&0.099629796142	&7998.8393&	903.49865\\
0.8879812490	&0.063719868179	&8035.5482&	576.61586\\
0.1459389534	&0.040532094854	&1317.0978&	365.80182\\
0.8054028192	&0.025777989322	&7244.4968&	231.86976\\
0.4484391912	&0.016297602508	&4016.6364&	145.97641\\
0.1117846249	&0.010312319717	&995.97978&	91.880810\\
0.3457884620	&0.006511918741	&3060.5873&	57.637249\\
0.0937131289	&0.004109395838	&822.62304&	36.072679\\
0.0472615616	&0.002596748497	&410.62217&	2.5613052\\
0.8301374652	&0.001638066550	&7120.4666&	14.050441\\
0.9363713955	&0.001034307919	&7905.0272&	8.7318261\\
0.9321481567	&0.000652563554	&7715.9636&	5.4016699\\
0.2871527054	&0.000411713254	&2320.0025&	3.3263687\\
0.4334166973	&0.000259844623	&3399.2927&	2.0379647\\
0.3973313201	&0.000163953500	&3005.5475&	1.2401993\\
0.8771240974	&0.000103425094	&6350.6295&	0.7488272\\

\hline
\end{tabular}
\end{table}

\begin{table}
    \centering
    \caption{Integrated $\nu_{\tau}$ downwards and upwards numbers at all zenith angles}
    \label{tab:your-label}
    \begin{tabular}{|c|c|c|}
    \hline

Total $\nu_{\tau}$Upwards	&Total $\nu_{\tau}$Downwards &$\theta$  \\
\hline
7185.48381746911	&116.084629999609&        54.365\\
6151.19253586331	&85.8860520803666&        48.369\\ 
5554.31120682837	&99.3200366451839&        42.697\\  
5095.4393091881	    &112.679601220319&        37.257\\   
4758.09970924391	&126.516716370702&        31.987\\  
51192.4720593863	&140.745478971196&        25.125\\  
40308.0573726772	&155.963585007827&        20.365\\
30855.1588364001	&172.544835989786&        14.695\\
29791.9258006516	&189.078054880136&        08.369\\

\hline
\end{tabular}
\end{table}

\begin{table}
    \centering
    \caption{Matter effects $\nu_{\mu}$--$\nu_{\tau}$ calculated variables for each bin defined in the paper}
    \label{tab:your-label}
    \begin{tabular}{|c|c|}
    \hline
A&C  \\
\hline
0.00495256258271	&1.00152364702271\\
0.006234922537161	&1.00192092281406\\
0.007871036961806	&1.00242942851609\\
0.009905125165419	&1.00306417197023\\
0.012469845074323	&1.00386850182387\\
0.015697854614839	&1.00488714527236\\
0.019766031022065	&1.00618081869592\\
0.024851251531097	&1.00781325834673\\
0.031307270612129	&1.00990994455869\\
0.039399404117807	&1.01257547411764\\
0.049614064444645	&1.01599821781034\\
0.062481883297936	&1.02039919721615\\
0.078621931000516	&1.02605393739413\\
0.099007032345419	&1.03339789534487\\
0.124610012125677	&1.04291670135913\\
0.156890107530839	&1.05533739157646\\
0.197527652294323	&1.07154169887191\\
0.24864517323729	&1.09263157605601\\
0.313028486812516	&1.11994022934473\\
0.394126699104387	&1.15480532955788\\

\hline
\end{tabular}
\end{table}

\section{Results}

The number of $\nu_{\tau}$ from oscillations of $\nu_{\mu}$, which we have calculated earlier from Bartol flux given in [2], is greatly affected by changes in baseline length and matter interaction, we must consider earth's matter density of $4.3 \, \text{gm/cm}^3$. We can plot the data from the file and perform fitting to see how the flux of $\nu_{\tau}$ changes at different zenith angles. Firstly, we want to see how the flux of $\nu_{\tau}$ changes with different energy scales. Secondly, we aim to calculate how latitude and longitude affect the flux. We start by integrating the Bartol flux to find the number of $\nu_{\tau}$ for different energies and specific zenith angles. After integrating, we check the variability of $\nu_{\tau}$ flux concerning different zenith angles and specific energy scales. The number of $\nu_{\tau}$ is directly proportional to the zenith angle at a fixed energy in $(\text{GeV})$ range and inversely proportional to energy at a fixed zenith angle $(\cos\theta)$. 

The variation of $\nu_{\tau}$ flux concerning zenith angle and energy is quite proportional but not a rapid change. The probability of getting $10,000 - 80,000$ $\nu_{\tau}$ is always there, irrespective of the zenith angle. This probability can oscillate into $\nu_{\tau}$ flux before interaction with the detector at INO and JUNO. 

The matter-induced effect on neutrino oscillation can be given by: 

\begin{equation}
\begin{aligned}
P_{\nu_{\mu}\rightarrow \nu_{\tau}} = \sin^2\theta_{m} \sin^2(\frac{\Delta m_{m}^2 L}{E})
\end{aligned}
\end{equation}

This is approximately equivalent to vacuum oscillations multiplied by 1.24:

\begin{equation}
\begin{aligned}
P_{\nu_{\mu}\rightarrow \nu_{\tau}} = 1.24 \times \sin^2\theta_{m} \sin^2(\frac{\Delta m_{m}^2 L}{E})
\end{aligned}
\end{equation}

When we include the cross-sectional area of the detector, the probability of the generation of an event will increase. The upward-going muon neutrinos have a higher baseline, hence the probability of conversion will also increase. The probability of conversion from muon to $\nu_{\tau}$ is approximately 0.69 compared to 0.05 when the baseline is taken to be 10 km for downward-going muon neutrinos.

\begin{equation}
\begin{aligned}
P_{\nu_{\mu}\rightarrow \nu_{\tau}} = 1.24 \times 0.8 \sin^2(\frac{0.001 \times 10000}{5})
\end{aligned}
\end{equation}

Also, the plots of flux vs. energy (GeV) provide comparisons and the flux vs. zenith angle is mentioned in Fig[3]. Where the mixing angle and masses of neutrinos will deviate from the standard by the following factors: The quasi-elastic scattering factor, which comes into play at this energy scale of neutrinos, is considered using the NUANCE event generator. Various large baseline experiments are analogous to short baseline experiments. Reactor $\nu_{e}$ disappearance experiments concerning atmospheric experiments, such as Kamiokande [81], IMB [82], Super-Kamiokande [3], Soudan-2 [83], MACRO [84], detect neutrinos that travel distances from about 20 km (downward-going) to about 12780 km (upward-going). They cover a wide energy spectrum from about 100 MeV to about 100 GeV (see Section 4.2). The probability depends upon the energy of atmospheric $\nu_{\mu}$. When we include the detector, the cross-section will increase because the density of interacted particles in materials is far larger than in the atmosphere. For $\nu_{\mu}$ to interact, the probability of quasi-elastic scattering will be higher when $\nu_{\mu}$ interacts with the detector medium. We will discuss the interaction in detail later, explaining how they interact with the detector material.
The plot of zenith angle and flux of $\nu_{\tau}$ and energy levels and flux of $\nu_{\tau}$.
The matter (MSW) affects the neutrino's oscillation with a given earth profile and density [6] as follows:

\begin{equation}
\begin{aligned}
\Delta m^2_{m} = C \times \Delta m^2 \hspace{1cm}
\sin^2\theta_{m} = C \times \sin^2\theta
\\
C = \sqrt{\cos^2(2\theta - A) + \sin^2 2\theta} \hspace{1cm}
A = \pm \frac{2\sqrt{2} G_f N_e E}{\Delta m^2}
\end{aligned}
\end{equation}

Including the number of $\nu_{\tau}$ that come after interaction with matter and energy (GeV) with the zenith angle, we plot the Bartol [2] flux with around 600 data sets in CSV format. The link will be mentioned. The probability that $\nu_{\tau}$ will be found at the INO site in India will be given concerning 600 data sets. 1D and 3D atmospheric neutrino fluxes use AGLS1996 fluxes energy and zenith angle fluxes in bins of neutrino energy (equally spaced bins in logE with 10 bins per decade with the low edge of the first bin at 100 MeV) and zenith angle (20 bins equally spaced in $\cos(\text{zenith})$ with bin width 0.1), integrated over azimuthal angle. LogE means to log to base e. Fluxes smaller than 10 GeV are from 3D calculations, and fluxes above 10 GeV are from 1D calculations. After analysis of data sets, one can find the various fluxes of $\nu_{\tau}$ having matter effects [1] at different zenith angles and other energies.

\begin{figure}[h]
    \centering
    \includegraphics[width=0.60\textwidth]{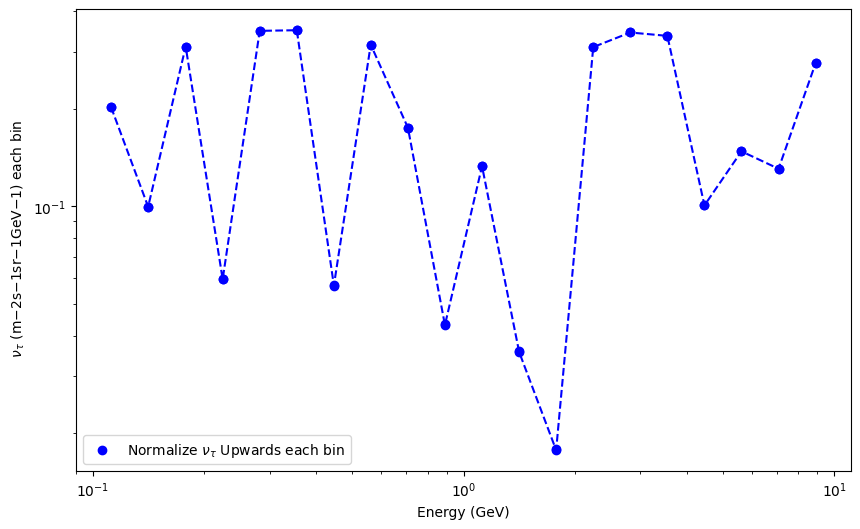}
    \caption{The plot predicts the $\nu_{\tau}$ number of events coming after matter interaction from the downside or the other side of the earth's atmosphere towards the detector to energy in GeV.}
\end{figure}

%\includegraphics[scale=0.75]{flux-energy.png} 
%\includegraphics[scale=0.55]{fit-Evsmuon.png} 

%When the energy of parent nuclei increases in the decay channel the probability of getting the muon number will reduce and go in favor of production of kaons hence the number decreases asymptotically and it goes to maximum as shown in in result section $T. Gaisser^4[2] $ after 100 GeV ranges, Also the branching ratio at high energy will favor the kaon production and not muon production in atmosphere hence the plot shows the co-relation as above.  
    \begin{figure}[h]
    \centering
    \includegraphics[width=0.60\textwidth]{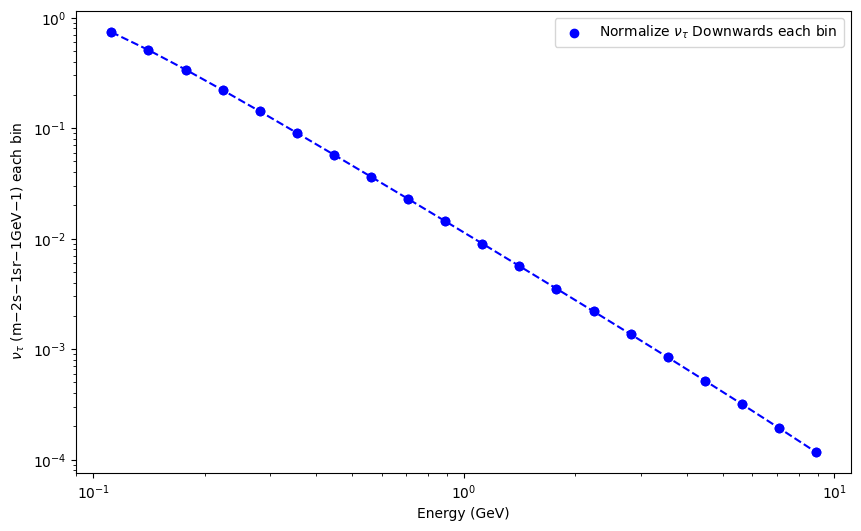}
    \caption
{
This plot shows the relationship between downward $\nu_{\tau}$ which increases when we move towards a higher zenith angle.
}
\end{figure}
\begin{figure}[h]
    \centering
    \includegraphics[width=0.60\textwidth]{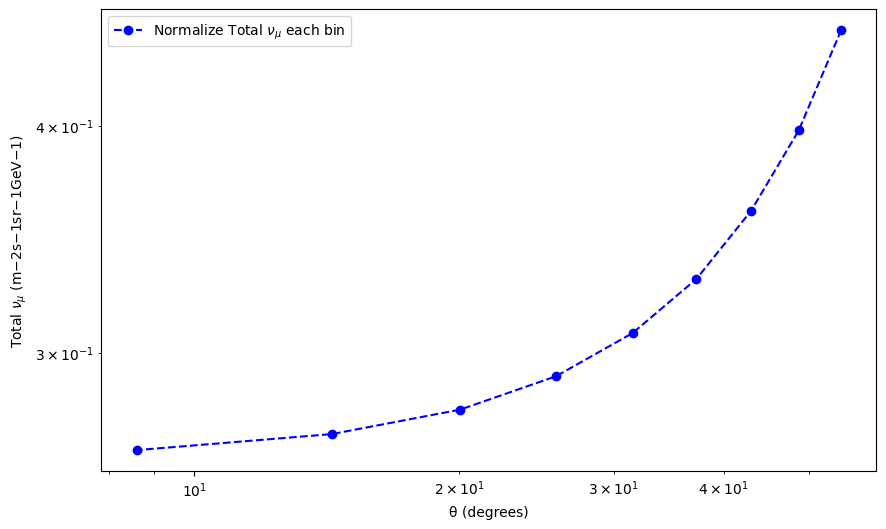}
    \caption
%includegraphics[scale=0.55]{fitt-ZvsMuon.png}
%\includegraphics[scale=0.55]{zenith-energy.png}
{
This plot represents the relationship between $\nu_{\mu}$ and the mentioned zenith angles.
}
\end{figure}

%Fractional contribution of pions and kaons to atmospheric muons and
%$\nu_{\mu}$. Solid lines are for vertical and dashed lines are for 60˝.[2]

%\includegraphics[scale=0.40]{flux-energy-zenith.png}

% Atmospheric neutrino fluxes (ν`Â¯ν) at zenith angles of 41˝(139˝)  and 76˝ (104˝). Points are from Ref. [2]. See the text for a discussion of the lines.%

\begin{figure}[h]
    \centering
    \includegraphics[width=0.60\textwidth]{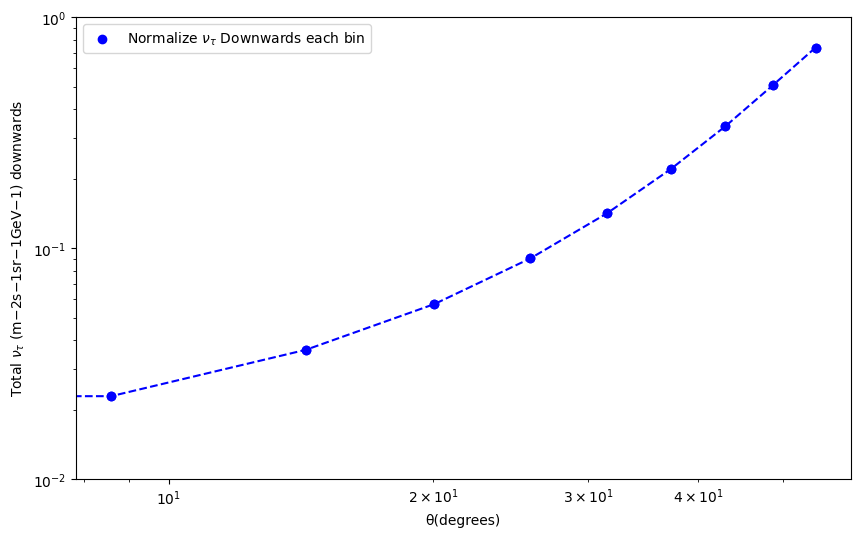}
    \caption
%includegraphics[scale=0.55]{fitt-ZvsMuon.png}
%\includegraphics[scale=0.55]{zenith-energy.png}
%\includegraphics[scale=0.45]{fit-Evstau-up.png}
{
The probability of getting the $\nu_{\tau}$ decreases as the $\nu_{\mu}$ decreases due to the decay channel production of $\mu$ concerning energy which has a significant effect on energy beyond 100 GeV hence the $\nu_{\tau}$ which is coming from above 30 km baselines shows asymptotic decreasing behaviour.
}
\end{figure}

\begin{figure}[h]
    \centering
    \includegraphics[width=0.60\textwidth]{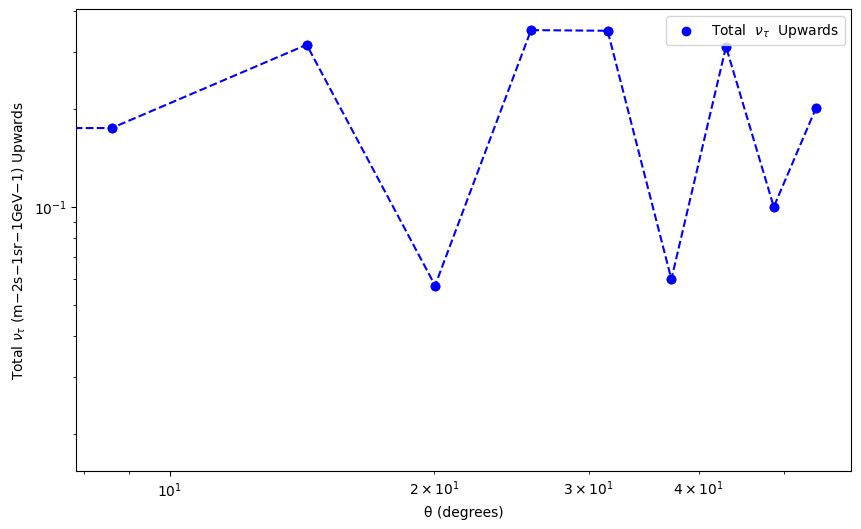}
    \caption
{
The plot shows the relationship between $\nu_{\tau}$ upwards to the mentioned zenith angles. 
}
\end{figure}
%\label{fig:mesh1}
%\includegraphics[scale=0.45]{fit-Zvstau-up.png}

%Zenith angle distributions of four classes of atmospheric neutrino events in Super-K (from a review article by Nakamura and Petcov in [10]). Dotted histograms show the neutrinos at production in the atmosphere from Monte Carlo simulation, and the solid histograms show the best fit for $\nu_{\mu}->\nu_{\tau}$ oscillations.

%[Thomas. gaisser]
%\setlength{\topmargin}{0in}
%\title{Blah Blah Cool}
%\date {}
$\\$
$\\$

In this case, we have a bigger baseline and a higher probability of detecting tau numbers, which are mentioned in the explanation section. We have also included the data sets calculated for the above calculations. The calculated fluxes for $\nu_{\tau}$ come from oscillations of $\nu_{\mu}$ and include matter effects of the earth's profile, named as the MSW data set, in a processed Excel data file while analyzing Bartol [3] flux data mentioned above in the earlier section. The tau numbers upwards are shown vs. energy in the GeV scale on a logarithmic scale to describe the clarity of data points and the fitting function, as shown in Plot I. The normalization factor is also calculated, and the actual number can be retrieved once the y-axis values are divided by the normalization factor of $4.3479\times10^{-5}$ [Fig 3].
Zenith angle distributions of four classes of atmospheric neutrino events in Super-K (from a review article by Nakamura and Petcov in [10]). Dotted histograms show the neutrinos at production in the atmosphere from Monte Carlo simulation, and the solid histograms show the best fit for $\nu_{\mu} \rightarrow \nu_{\tau}$ oscillations.
The atmospheric secondary flux contribution is compared in Figure 8 with that of the primary component. At 0.1 GeV, nearly 50 per cent of the neutrinos come from secondary proton/neutron-induced reactions. This fraction decreases to about 20 per cent at the 50 GeV energy scale.
Results show that there is a high probability that $\nu_{\tau}$ event generation is quite high when conducting an actual experiment at INO India. With careful and proper handling of sampling and processing of signals, which are crucial in detection mechanics and instruments involved, our calculations provide evidence that the probability of obtaining $\nu_{\tau}$ from oscillations, including matter effects in the earth's density profile, is high due to the large baseline at different zenith angles. We can ignore the earth's geomagnetic effects on muon and $\nu_{\tau}$ involved in decay channels and oscillations.
When we increase the zenith angle, it will also increase the volume of the cone defined concerning the zenith angle. Hence, the probability of events occurring in that cone increases due to the larger volume as the zenith increases. Thus, the production of downward $\nu_{\tau}$ will increase when we move towards a higher zenith angle. The normalization factor will give the actual $\nu_{\tau}$ number as discussed earlier, which is equal to $1.565\times10^{-4}$ [Fig 4].
The plot represents the possible number of $\nu_{\mu}$ at different zenith angles. These numbers are predicted and measured by the Bartol team and analyzed from the data sets mentioned in [14]. The reason for this plot is to show how the muon number increases significantly when the zenith angle increases. As the zenith angle increases, the surface area and volume of the cone also increase. This larger volume includes a higher probability of securing more events in that particular region. Hence, the enhanced cone region due to the increment of the zenith angle increases the probability of an event happening in that region. This is shown in the plot. The normalization factor is $2.9950\times10^{-6}$[Fig 5].
We can also include the elastic and quasi-elastic scattering and other deep inelastic scattering and interaction of neutrinos with the detector, specifically 10 kilotons of $Iron (Fe)$ nuclei, depending upon its energy of interaction at INO India. We can provide detailed explanations of interactions and results later with the relevant analysis of required data sets. 
The probability of getting $\nu_{\tau}$ upwards will greatly increase after matter interaction from the downside of the earth to be detected with a density of $10.3gm/cm^3$ $\nu_{\tau}$ will show somewhat sensual behaviour when calculated from muon numbers of $Bartol^3[3]$ flux even when fitted polynomial fitting parameter marginally changes and shows linearity in fitting data points. We can get the approximate numbers of $\nu_{\tau}$ after considering the normalization factor with simple calculations from the y-axis values shown and having a normalised factor $4.3479\times10^{-5}$ [Fig 7] which gives the actual number when divided. 
The interaction and specifications of the detector will be discussed later, detailing how interaction with neutrinos with the sensor affects the probability of $\nu_{\tau}$ events. Generally, for a large baseline, the ratio of (L/E) will affect the probability of oscillations for $\nu_{\tau}$. A small and fixed baseline energy $(E > 1 \, \text{TeV})$ will decrease the $\nu_{\tau}$ events because of probability oscillations effects and other detector specifications, as mentioned in the results section [B].
Investigation of various data sets needs to be collected for detailed analysis of latitude and longitudinal effects on $\nu_{\tau}$ flux at INO India for different zenith and azimuthal angles. Their detection and generation of events will be discussed in the next paper.

\section{Summary}

The study of decay channels with proper resolution helps in differentiating whether it is a $\nu_{\tau}$ particle with corresponding leptons compared with background $\nu_{\mu}$, with scrutinized direction as well. A waveform of hits has been used to record, and detector efficiency has been taken care of in the case of JUNO and INO. Integrating the Bartol flux for all zenith and all energy bins has given the muon numbers and has taken care of matter profile interaction of $\nu_{\mu}$, which leads to oscillations from muons to $\nu_{\tau}$. $\nu_{\tau}$ is a rare atmospheric event, and we are studying the different channels. For example, pions and kaons decay channels lead to the generation of $\nu_{\mu}$ fluxes at all zenith angles. Depending upon their energies, they may undergo further decay and oscillate to different flavours as well. Hence, we are more interested in muon neutrino flavour conversion to $\nu_{\tau}$ as a possible source in the atmosphere. Zenith and azimuthal angle distribution and flavour ratios at different latitudes, corresponding energy distributions, production altitude, and production angle distributions have been calculated and discussed for neutrino flux. The components originating from cosmic rays and the atmospheric cascade have been distinguished and discussed separately. Calculated flux observed features could be traced back to the primary spectra, $(K^+/K^-)$ ratio, and $(\pi^+/\pi^-)$ ratios, $\mu$ kinematics, $\pi$ kinematics, geomagnetic latitude and longitudinal effects, and geometry. 
$\\$
The paper's primary focus is to provide precise calculations of the expected $\nu_{\tau}$ flux, which is dependent upon energy and zenith angle, considering many factors, including atmospheric composition, neutrino oscillations, and the energy spectrum of cosmic rays. These calculations are based on state-of-the-art simulation techniques and utilize the latest data on Super-K Japan. One of the critical aspects of the research is its relevance to neutrino observatories, such as INO India and Super-K in Japan, which aim to detect low to high-energy neutrinos. Our study specifically comprised the energy ranges of 1-10 GeV. The precise predictions for $\nu_{\tau}$ flux presented in this paper serve as essential inputs for designing and analysing experiments at these observatories. Specific 3-dimensional effects seen in the simulation results are minimal, which indicates that the 1-dimensional approximation provides reliable results for GeV neutrino flux calculations. The reproduced calculations do not quantitatively include the east-west asymmetry of atmospheric neutrino-induced events in the Soudan, Kamioka, and Super-K detectors. Further investigations are needed to understand this disagreement. For example, as mentioned in [14], there are possibilities of different fluxes for $\cos(-0.95)$ to $\cos(0)$ and then $\cos(0)$ to $\cos(+0.95)$ angles, as experimentally deduced in different experiments mentioned. and this can be true for all the specific zenith angles estimation for all the energy of $\nu_{\tau}$ downward and upgoing events after interaction with the detector to produce preliminary data.

% Acknowledgements
\acknowledgments
Special gratitude to the EHEP Lab at IISER Mohali for providing the necessary ambience and facilities to complete this project.


\begin{thebibliography}{99}

\bibitem{1}
K. Nakamura and S.T. Petcov,
"Neutrino Mass, Mixing, and Oscillations,"
\emph{J. Abbrev.}, vol. [Volume], May 2014, pp. [Pages].

\bibitem{2}
T.K. Gaisser, T. Stanev, S.A. Bludman, and H.S. Lee,
"Flux of Atmospheric Neutrinos,"
\emph{Phys. Rev. Lett.}, vol. 51, 1983, p. 223.

\bibitem{3}
G.D. Barr, T.K. Gaisser, P. Lipari, S. Robbins, and T. Stanev,
"A Three-Dimensional Calculation of Atmospheric Neutrinos,"
\emph{Phys. Rev. D}, vol. 70, 2004, p. 023006.

\bibitem{4}
A. Yu. Smirnov,
"Flux of Atmospheric Neutrinos,"
\emph{Phys. Scr.}, vol. T121, 2005, pp. 57–64.

\bibitem{5}
T.K. Gaisser and M. Honda,
"The MSW Effect and Matter Effects in Neutrino Oscillations,"
\emph{Annu. Rev. Nucl. Part. Sci.}, vol. 52, 2002, pp. 153–99.

\bibitem{6}
Q. Gani and W. Bari,
"Neutrino Interaction Cross-sections from MeV to GeV Scales of Energy,"
\emph{SSRG Int. J. Appl. Phys.}, vol. 4, no. 2, Sep.-Dec. 2017.

\bibitem{7}
T.K. Gaisser, R. Engel, and E. Resconi,
\emph{Cosmic Rays and Particle Physics},
University of Delaware, Karlsruhe Institute of Technology, Technical University Munich.

\bibitem{8}
"Secondary Cosmic Rays,"
\emph{ScienceDirect}.
\url{https://www.sciencedirect.com/topics/physics-and-astronomy/secondary-cosmic-rays}

\bibitem{9}
R.T. Senthil, D. Indumathi, and P. Shukla,
"A Simulation Study of $\nu_{\tau}$ Events at the ICAL-INO,"
\emph{Homi Bhabha National Institute}, Mumbai.

\bibitem{10}
\textcolor{blue}{\emph{https://www.ino.tifr.res.in/ino/index.php}}

\bibitem{11}
M. He, Z. Sun, L. Wan, and X. Zhang,
"JUNO physics and detector,"
\emph{Progress in Particle and Nuclear Physics}, vol. 123, p. 103927, 2022.
\url{https://doi.org/10.1016/j.ppnp.2022.103927}

\bibitem{12}
S. A. Meighen-Berger, J. F. Beacom, N. F. Bell, and M. J. Dolan,
"New Signal of Atmospheric $\nu_{\tau}$ Appearance: Sub-GeV Neutral-Current Interactions in JUNO,"
2023.
\url{https://arxiv.org/pdf/2311.01667.pdf}

\bibitem{13}
C. A. Argüelles, M. Bustamante, A. Connolly, and C. Finley,
"$\nu_{\tau}$ in the next decade: from GeV to EeV,"
\emph{Journal of Physics G: Nuclear and Particle Physics}, vol. 49, no. 10, 2022.
\url{https://doi.org/10.1088/1361-6471/ac89d2}

\bibitem{14}
G. D. Barr,
"Neutrino flux predictions and estimation,"
\emph{Flux files for neutrino experiments},
2004.
\url{http://www.pnp.physics.ox.ac.uk/~barr/fluxfiles/0401/fmin10_0401z.kam_num}

\bibitem{15}
C. Giunti and C. W. Kim,
"Introduction to neutrino physics,"
in \emph{Lectures on Astroparticle Physics}, 2004.
\url{https://arxiv.org/pdf/hep-ph/0408165}

\bibitem{16}
G. P. Lepage,
"Introduction to Monte Carlo methods,"
2000.
\url{https://arxiv.org/abs/hep-ph/0006269}

\bibitem{17}
India-based Neutrino Observatory (INO), 
"INO homepage,"
2023.
\url{https://www.ino.tifr.res.in/ino/index.php}


\end{thebibliography}
\end{document}